\documentclass[10pt, preprint]{aastex}
\usepackage{epsfig}

\shorttitle{Multi-Object AO Demonstrations}
\shortauthors{Ammons et al.}

\begin{document}

\title{Integrated Laboratory Demonstrations of Multi-Object Adaptive Optics on a Simulated 10-Meter Telescope at Visible Wavelengths} 

\author{S. Mark Ammons\altaffilmark{1}, Luke Johnson\altaffilmark{1}, Edward A. Laag\altaffilmark{2}, Renate Kupke\altaffilmark{1}, Donald T. Gavel\altaffilmark{1}, Brian J. Bauman\altaffilmark{3}, Claire E. Max\altaffilmark{1}}

\altaffiltext{1}{present address:  Center for Adaptive Optics, University of California, Santa Cruz, 1156 High St., Santa Cruz, CA  95064, ammons@ucolick.org, ljohnson@ucolick.org, kupke@ucolick.org, gavel@ucolick.org, max@ucolick.org}
\altaffiltext{2}{present address:  University of California, Riverside, 900 University Ave., Riverside, CA USA  92521, laag@ucolick.org}
\altaffiltext{3}{present address:  Lawrence Livermore National Laboratory, 7000 East Avenue, Livermore, CA, 94550, USA, bauman3@llnl.gov}

\begin{abstract}
One important frontier for astronomical adaptive optics (AO) involves methods such as Multi-Object AO and Multi-Conjugate AO that have the potential to give a significantly larger field of view than conventional AO techniques.  A second key emphasis over the next decade will be to push astronomical AO to visible wavelengths.  We have conducted the first laboratory simulations of wide-field, laser guide star adaptive optics at visible wavelengths on a 10-meter-class telescope.  These experiments, utilizing the UCO/Lick Observatory's Multi-Object / Laser Tomography Adaptive Optics (MOAO/LTAO) testbed, demonstrate new techniques in wavefront sensing and control that are crucial to future on-sky MOAO systems.  We (1) test and confirm the feasibility of highly accurate atmospheric tomography with laser guide stars, (2) demonstrate key innovations allowing open-loop operation of Shack-Hartmann wavefront sensors (with errors of $\sim30$ nm) as will be needed for MOAO, and (3) build a complete error budget model describing system performance.    The AO system maintains a performance of 32.4\% Strehl on-axis, with 24.5\% and 22.6\% at $10\farcs$ and $15\farcs,$ respectively, at a science wavelength of 710 nm (R-band) over the equivalent of 0.8 seconds of simulation.  The mean ensquared energy on-axis in a 50 milliarcsecond spaxel is 46\%.  The off-axis Strehls are obtained at radial separations 2-3 times the isoplanatic angle of the atmosphere at 710 nm.  The MOAO-corrected field of view is $\sim25$ times larger in area than that limited by anisoplanatism at R-band.  The error budget we assemble is composed almost entirely of terms verified through independent, empirical experiments, with minimal parameterization of theoretical models.  We find that error terms arising from calibration inaccuracies and optical drift are comparable in magnitude to traditional terms like fitting error and tomographic error.  This makes a strong case for implementing additional calibration facilities in future AO systems, including accelerometers on powered optics, 3D turbulators, telescope and LGS simulators, and external calibration ports for deformable mirrors.   These laboratory demonstrations add strong credibility to the implementation of on-sky demonstrators of Laser Tomographic Adaptive Optics (LTAO) on 5-10 meter telescopes in the coming years.
\end{abstract}


\section{INTRODUCTION}

\subsection{Progression of Adaptive Optics Technologies}

The advent of laser guide star (LGS) adaptive optics (AO) on 8-10 meter class telescopes has enabled a new regime of diffraction-limited science at near-infrared wavelengths.   Adaptive optics uses deformable mirrors to rapidly ($\sim1$ kHz) correct atmospheric-turbulence-induced phase errors.  This process requires a bright, point-like phase reference, a luxury that is quite rare among the stars in the sky.  In the 1980's and 1990's, Lawrence Livermore National Laboratory (LLNL) pioneered use of the sodium-layer laser guide star to increase the sky visibility of AO systems.  In LGS adaptive optics (LGS-AO), a laser is tuned to the sodium D2 line (589 nm) and focused on the atmosphere's sodium layer at 90 km, creating an ``artificial star'' in the sky \citep{foy85, max97}.   

The laser guide star has since contributed greatly to high spatial resolution astronomy at infrared (IR) wavelengths.  Most 8-10 meter class telescopes now have a LGS-AO program, and many exist on smaller telescopes.  The LGS facility at Gemini North has been routinely used since February 2007 \citep{tru07} to study the stellar populations of nuclear clusters in nearby galaxies \citep{set08} and resolve distant quasar hosts \citep{wat08}, for example.  The Very Large Telescope's (VLT) LGS system has been used with a near-infrared imager and intregal field unit (IFU) spectrograph since 2007 \citep{bon06} to investigate the kinematics of high-redshift galaxies \citep{gen08} and resolve gravitationally-lensed quasars \citep{slu08}.  The LGS-AO system at the Keck Observatory \citep{wiz06} is extremely productive scientifically, having been used to directly image extrasolar planets \citep{mar08}, track the motion of stars orbiting the Milky Way Galaxy's central black hole \citep{ghe05}, localize the progenitors of Type II supernovae \citep{gal07}, and investigate the binarity of brown dwarfs \citep{liu06}.  Including such systems on all telescopes, 60 refereed journal articles based on LGS-AO imaging and spectroscopy have been published \citep{liu08}.  

\subsection{Rationale for Visible- and Wide-Field AO}

The success of the current generation of LGS-AO systems in terms of peer-reviewed scientific articles is encouragement to push adaptive optics technologies further.

Space-based facilities such as the superb 2.4 meter Hubble Space Telescope (HST) routinely deliver field sizes of 2-4 arcminutes for optical/near-IR imaging with a stable point spread function (PSF).  HST will be refurbished in 2009 with the Wide Field Camera 3 (WFC3) as well as other new instruments, giving images with full width at half max (FWHM) between 35 and 150 millarcseconds from the optical to H-band.  The James Webb Space Telescope (JWST) will deliver diffraction-limited imaging as well as integral-field unit (IFU) spectroscopy at wavelengths of 2 microns or longer with a 6.5 meter aperture.  Despite the unprecedented optical/IR capabilities of these facilities, the availability of even larger ground-based primary mirrors is an attractive reason to pursue diffraction-limited imaging from the ground.   The B and V passbands, which will not be imaged by JWST, are especially critical.  

Wide-field, laser-driven visible light adaptive optics on 5-10 meter telescopes will be an exciting new capability.   An 8-10 meter telescope with 30\% Strehl in V would be several times more sensitive to point sources in the visible than any telescope existing today.  This sensitivity would enable deeper studies of extragalactic globular clusters, supernovae, and quasars at high redshift, as well as bright OB stars, OII regions, and red giants in galaxies as distant as 30-100 Mpc.  Laser guide stars will be critical to these systems to enable good sky coverage.  The great potential of large-aperture visible-light AO has been explored in various science case studies for future AO systems, including the Keck Next Generation AO System \citep{max08} and PALM-3000 for the Palomar telescope \citep{bou08}.

\subsection{Laser Tomographic Adaptive Optics}

The performance of all AO systems is limited by errors in wavefront estimation and phase correction.  The ``cone effect'' is one such error inherent to the current generation of LGS-AO systems.  This error term is present because the cone of atmosphere sampled by a laser guide star does not overlap sufficiently with the cylinder of turbulence sampled by astronomical objects.  The cone effect error term increases with aperture size, reaching $\sim150$ nm RMS on a 10-meter aperture, which precludes laser-driven, diffraction-limited performance at visible wavelengths on 8-10 meter telescopes if single laser beacons are used.  

\begin{figure}
    \begin{center}
     \includegraphics[width = 6.0in, height = 2.2in]{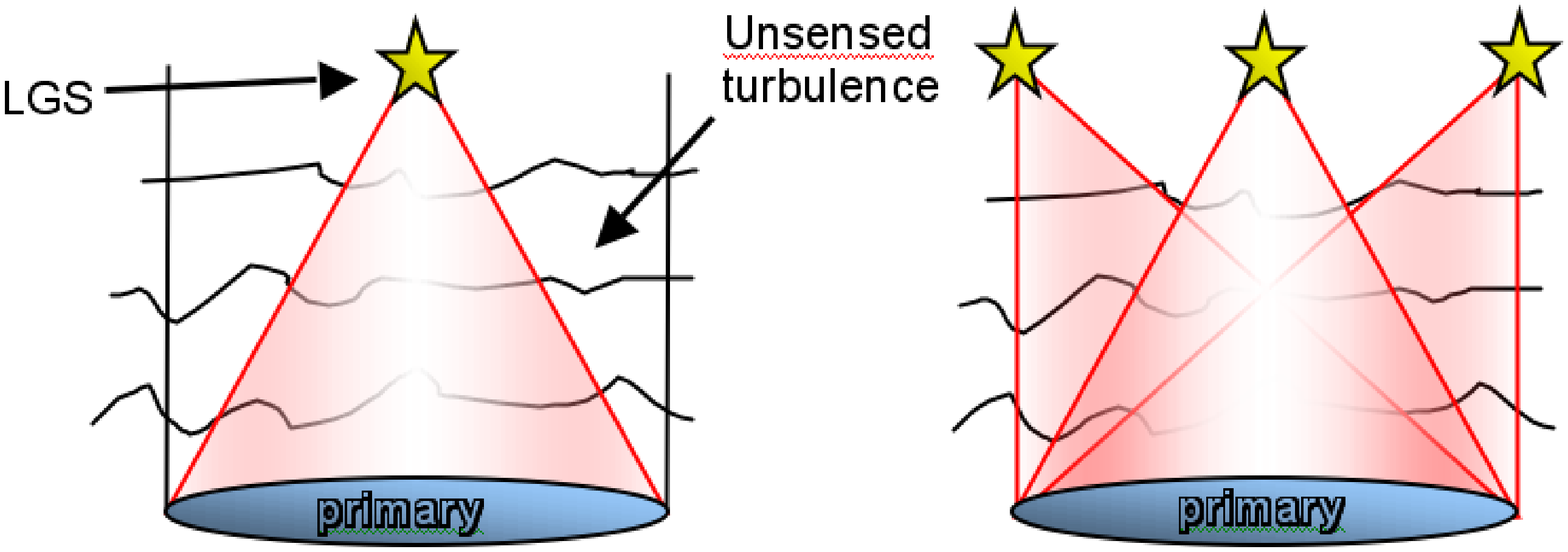}
       \caption{Multiple laser guide stars can sense turbulence outside the laser guide star cone, reducing the ``cone effect'' error term.  Left panel:  Single laser guide star case.  Right panel:  Multiple laser guide star case.}
   \end{center}
\end{figure}

To address this, multiple laser guide stars are envisioned (see Figure 1) to provide wavefront sensing of the entire cylinder of turbulence leading to the science object \citep{rag99, via00, tok01}.  In such an arrangement, the integrated LGS wavefronts can be analyzed tomographically \citep{gav04, bar06, gil08} to compute turbulence phase maps at various heights in the atmosphere.  The deformable mirror correction can then be optimized along a single line of sight toward an object of interest through the 3-D atmosphere.  AO systems that use laser tomography adaptive optics (LTAO) and solve the cone effect may reach higher Strehl on-axis than traditional LGS-AO systems of equivalent subaperture size, correction rate, and total laser brightness.

To optimize LTAO system performance on-axis, the LGS constellation size is best set to geometrically fill the cylinder of turbulence sampled by an astronomical object (see the left panel of Figure 2).  This arrangement minimizes the error in tomographic wavefront estimation.  

\subsection{Multi-Object Adaptive Optics}

Multi-Object AO (MOAO) is a wider-field variant of LTAO in which multiple scientific sensors, either spectrographs or small imagers, probe the large ``field of regard'' sensed by the laser guide stars \citep{ham04, ass04}.   Each of these sensors has a separate internal deformable mirror, applying a correction optimized for the direction in which it is pointed.  MOAO systems require multiple sensors and deformable mirrors to take advantage of the large field of regard sensed by the laser guide stars (typically $2\farcm$ to $5\farcm$).  

Each of the MOAO ``arms'' utilize three-dimensional wavefront sensing information from the laser guide stars, removing the cone effect and realizing the theoretically higher Strehl of LTAO than traditional single-conjugate LGS-AO.  In MOAO, the LGS constellation may be widened beyond the optimal LTAO radius to increase the field size.  However, the error in tomographic wavefront sensing increases with constellation radius, so the LTAO Strehl represents the upper limit to the MOAO system performance.   Figure 2 sketches the major differences between the high-Strehl LTAO and MOAO system designs.

LTAO and MOAO are currently being considered as pathways towards laser-driven, wide-field, visible light AO on large aperture telescopes.  These techniques must be tested with integrated testbeds to ensure their feasibility.

\begin{figure}
    \begin{center}
     \includegraphics[width = 6.0in, height = 4.2in]{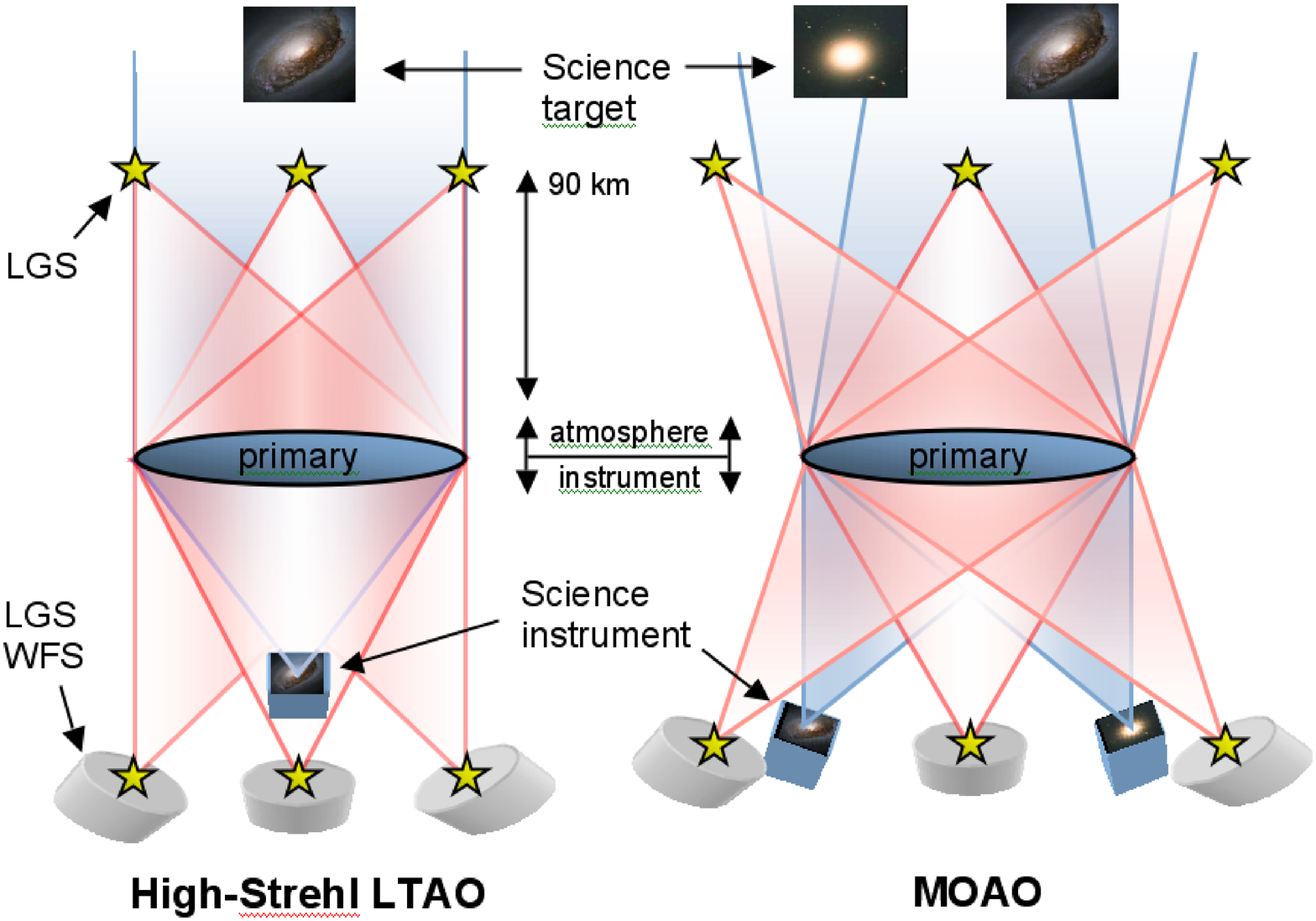}
       \caption{Simplified schematic diagrams of high-Strehl LTAO and MOAO architectures.  The left panel displays an LTAO design and the right panel shows an MOAO design.  LGS wavefront sensors are shown as gray cylinders; science instruments and integrated deformable mirrors are shown as blue boxes.  Folding optics and other optical relays are omitted.  If just one science instrument is used in LTAO mode (as shown), the LGS constellation size can be set to fill the cylinder of turbulence illuminated by an object at infinity, optimizing on-axis Strehl.  In MOAO, this constellation size is widened to increase the field of regard.  The field area in-between guide stars can be populated with the MOAO arms assigned to individual astronomical objects.}
   \end{center}
\end{figure}

\subsection{Goals of this Paper}

We have implemented an integrated MOAO/LTAO laboratory instrument in the Laboratory for Adaptive Optics at the University of California, Santa Cruz, and now rigorously test its performance at R-band.  The goals of this paper are to:
\begin{enumerate}
\item[(1)]Demonstrate the technical feasibility of the Laser Tomography AO and Multi-Object AO modes;
\item[(2)]Show that wavefront sensors and deformable mirrors with 64x64 degrees of freedom are of sufficient order to extend MOAO correction into visible wavelengths, in the limit of bright laser and tip/tilt guide stars;
\item[(3)]Demonstrate open-loop wavefront sensing with calibrated Shack-Hartmann wavefront sensors; and
\item[(4)]Construct a time-dependent error budget that models both the spatial and temporal performance of the system.
\end{enumerate}
Section 2 describes the optical layout of the testbed and includes an explanation of a calibration that enables open-loop wavefront sensor (WFS) operation.  The performance of the AO system for a $45\farcs0$ diameter LGS constellation and an optical science wavelength (710 nm) is presented in Section 3.  Section 4 assembles an error budget modeling total system performance, which is largely composed of empirically-measured terms with some parameterization of theoretical dependencies.  Section 5 concludes and discusses implications for MOAO/LTAO systems now being designed.

\section{UCO/LICK MOAO/LTAO TESTBED}

\subsection{Summary}

The UCO/Lick MOAO/LTAO testbed is an integrated laboratory simulator of a wide-field adaptive optics system on a 10-30 meter telescope, with optical layout shown in Figure 3.  The optical design of the testbed is summarized here and presented in more detail in sections 2.2 - 2.9.  Further details of the optical layout are also presented in earlier publications \citep{gav06, amm06, amm08}.

The system has five star-oriented Shack-Hartmann wavefront sensors pointing in the direction of five laser guide stars.  These wavefront sensors are set to 66x66 subapertures to enable high-order compensation.  The design of the wavefront sensors and a description of their open-loop performance are presented in sections 2.2-2.3.  The five laser guide stars are arranged in a box-5, or ``quincunx,'' constellation at an equivalent altitude of 90 km with a guide star diameter of $45\farcs0.$  Sections 2.4-2.5 describe the LGS constellation and its mechanical elongation.  The atmospheric section above the telescope aperture contains 3-5 Galil motors for simple translation of atmospheric phase plates.  The phase plates are glass, acrylic, or plastic slides with Kolmogorov turbulence etched into or sprayed onto them.  Further details on the phase plates are in Section 2.6.  The testbed has three deformable mirrors, conjugated to different altitudinal layers.  An ALPAO DM-52 woofer deformable mirror is conjugated to the ground layer, as is one Hamamatsu Programmable Phase Modulator (PPM).   A second PPM is conjugated to 9 km altitude.    The deformable mirrors are described in Section 2.8.  

The optical layout and performance of the MOAO/LTAO testbed are intended to be analogous to a wide-field laser guide star adaptive optics system on a 10-meter telescope.  This is accomplished by matching wavefront sensor type, subaperture number, the geometry of atmospheric metapupils, and dimensionless atmospheric strength to the values expected for such an on-sky system.  This matching of ``similarity parameters'' is discussed in Section 2.10.

\begin{figure}
   \begin{center}
   \begin{tabular}{c}
   \includegraphics[height=18cm]{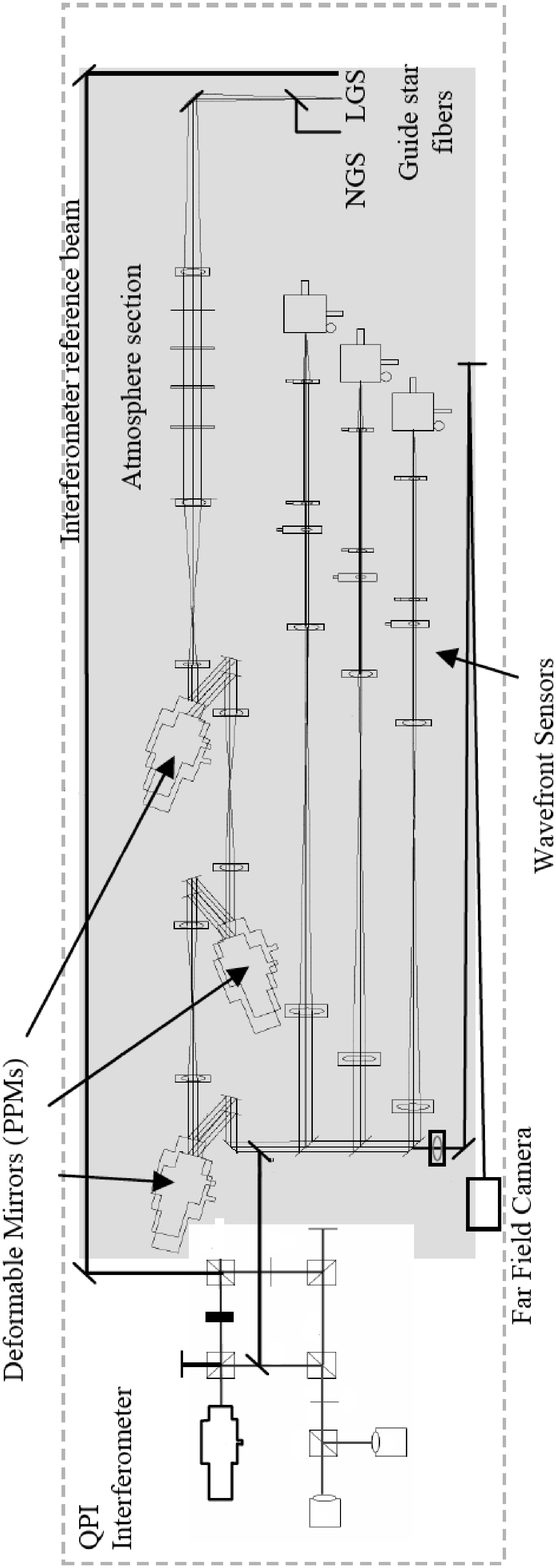}
   \end{tabular}
   \end{center}
   \caption[example] 
{ \label{fig:layout} 
Optical layout for MOAO/LTAO testbed.  The testbed is divided into four large sections by broad functionality:  An atmospheric section, a deformable mirror section, a wavefront sensor array, and a quadrature-polarization truth interferometer.  Five laser guide stars at 90 km equivalent altitude and seven science stars at infinity illuminate an atmospheric section with three translating atmospheric phase plates.  The light then passes through three ports where deformable mirrors can be conjugated to different altitudinal heights.  After exiting the deformable mirror section, the light is sent into one of several multiplexed wavefront sensors, which can detect up to nine laser guide stars.  The science light is sent to a far-field camera for Strehl checking.  The Quadrature Polarization Interferometer (QPI) is used to check the measurements of the Shack-Hartmann wavefront sensors as well as to measure the total magnitude of super-Nyquist frequencies.}
   \end{figure} 

\subsection{Multiplexed Wavefront Sensors}
The Shack-Hartmann wavefront sensor pupils are multiplexed onto two Dalsa cameras, allowing for up to eight laser guide stars with up to 100x100 subapertures per pupil, suitable for simulating MOAO on a 30-meter telescope.  The WFS multiplexing, designed by B. Bauman, is accomplished with two pupil-steering quad-lenses, with four off-axis sections of lenses glued together in each quad.  This structure images each LGS pupil onto a particular quadrant of the camera, regardless of the field position of the LGS.  Due to features on the cut edges of the quad-lenses, the laser guide stars cannot be placed closer than $10\farcs$ from each other.  The multiplexing design saves camera space and has been shown to be feasible on other on-sky systems (ViLLaGES, e.g., \citet{gav08})

Each Shack-Hartmann subaperture has 4x4 pixels for centroiding.  A Vitrum lenslet array with $f = 6.8$ mm is used to break up the pupil beam into subapertures, giving a Hartmann spot FWHM of 0.61 pixels.  This small spot size is chosen to minimize spot interference, but Hartmann linearity problems become significant, as described in section 2.3. 

\subsection{Open-Loop Performance of Wavefront Sensors}
For some architectures of Laser Tomographic AO systems, in which the wavefront sensing unit is not placed downstream of the deformable mirror(s), the wavefront sensors are required to operate in open-loop.  Open-loop wavefront sensor operation is particularly critical to MOAO systems.  They must sense the phase in a particular direction to high accuracy with high dynamic range, which are not properties typically endemic to Shack-Hartmann wavefront sensors.  The low dynamic range of the Shack-Hartmann WFS is caused by non-linearities inherent to centroiding unresolved structures with finite windows.  This effect is exacerbated in systems with small Hartmann spot sizes, especially less than one pixel in FWHM.  The various difficulties associated with open-loop wavefront sensing are described in more detail in \citet{amm07}.

We address this problem by calibrating the response of every subaperture in all pupils to linear signals.  These input linear signals are generated by raster scanning the laser guide star constellation in a box pattern, creating perfect tilts.  The Hartmann signals (x,y) are recorded by subaperture for each true tilt in [xt,yt] and inverted using a two-dimensional fit to (x,y) for both xt and yt, creating two linearized lookup tables in x and y for each subaperture.  This procedure has no direct analogy for an LTAO system on a telescope, as no internal calibrator could reproduce the size and shape of the real LGS spots in real time.  However, it may be possible to dither the actual laser guide star constellation during real-time operation.
 
In the MOAO/LTAO bench, the open-loop performance of the wavefront sensors can be checked by comparing the agreement of different laser guide star wavefronts through a common ground layer of turbulence.  These tests on our system suggest that the remaining error in open-loop wavefront measurements is $\sim30$ nm RMS for piston/tip/tilt-removed atmospheres of $600$ nm RMS.  This open-loop error is found to be proportional to the total amplitude of atmosphere introduced.  For the visible-light AO experiments presented in this paper, this systematic error averages $39$ nm.

\subsection{Guide and Science Stars}
High-power LEDs at $\lambda = 650$ nm are used for laser guide stars, at an equivalent altitude of $90$ km.  These LEDs have a nonzero spectral width of $\sim10-20$ nm, which decreases coherent interference between guide stars.  We use five LEDs in a ``quincunx'' pattern of $22\farcs5$ radius.  Seven science test stars are distributed within the LGS constellation, with off-axis angles up to $15\farcs0$.  The light from these natural test stars is collimated at the telescope pupil.  The constellation is shown in Figure 4.  The wavelength of the science stars is $658$ nm.  The error due to the difference in central wavelengths between the wavefront sensing stars and the science test stars is less than 5 nm RMS.   In addition, the multispectral error due to the LED's nonzero spectral width (arising from nonlinear dependencies between effective PPM phase and wavelength) is less than 2 nm RMS.

This AO system has multiple science stars so that the Strehl can be checked in different directions.  A conventional MOAO system design has multiple sensors and multiple deformable mirrors to realize the highest Strehls across the field of regard simultaneously.  In this testbed, however, there is only one ground-layer PPM deformable mirror.  We simulate conventional MOAO by checking Strehl on different science stars while the atmosphere is ``frozen'' during a single AO iteration.  This involves optimizing the correction for 7 different field points and applying 7 different corrections with the ground-layer PPM during an AO iteration.  Because the atmosphere is motionless during this process, this simulates simultaneous correction with multiple deformable mirrors in different directions.  A two-phase paddle blocks either the laser guide stars during Strehl measurement or the science stars during wavefront sensing.

\begin{figure}
   \begin{center}
   \begin{tabular}{c}
   \includegraphics[height=10cm]{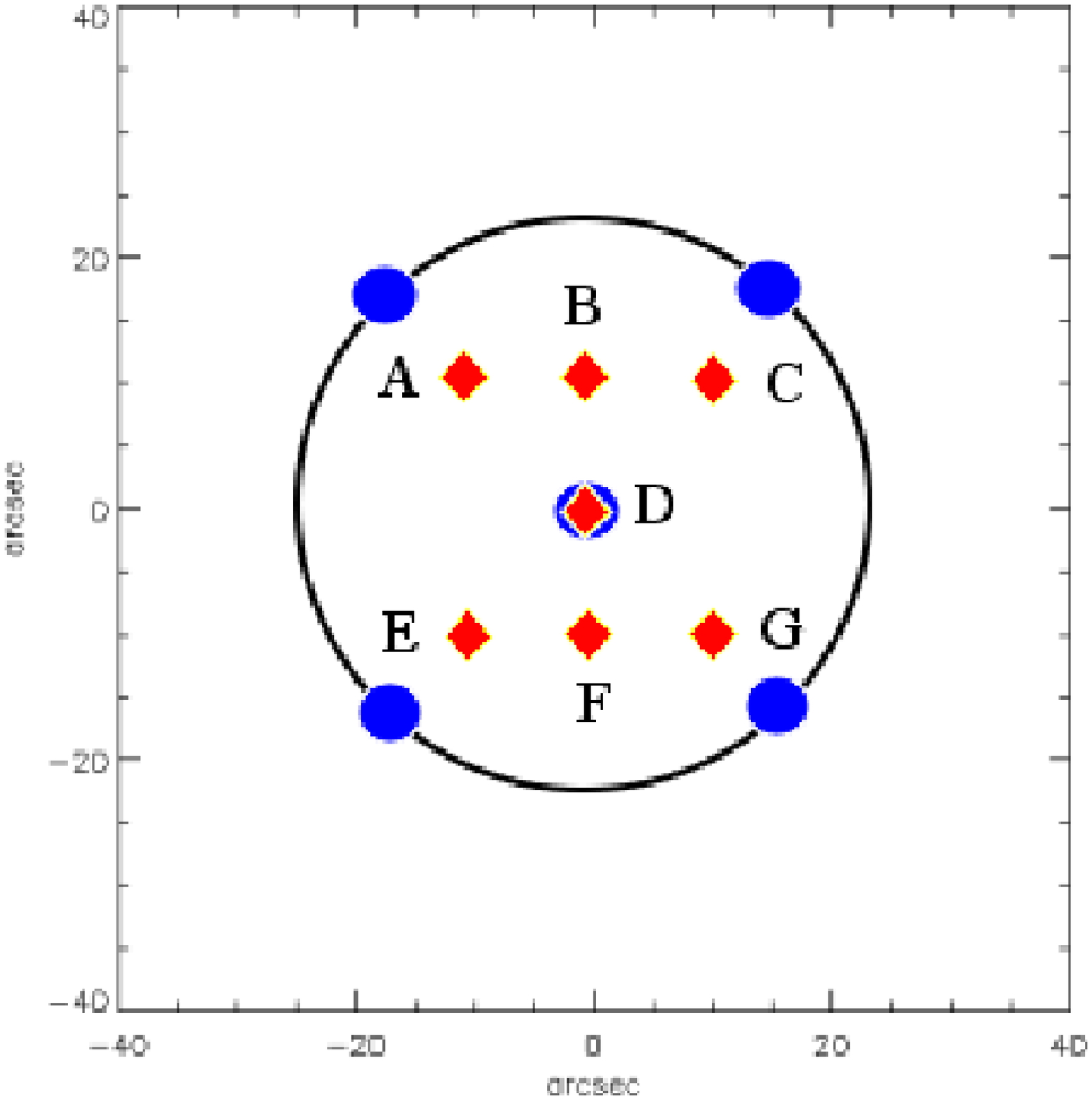}
   \end{tabular}
   \end{center}
   \caption[example] 
{ \label{fig:constellation} 
LGS and science star constellation for MOAO experiments.  Blue circles denote the positions of laser guide stars; red diamonds denote the positions of science test stars.}
   \end{figure} 

\subsection{LGS Elongation}

The laser guide star constellation is mechanically elongated by physically moving the LGS constellation by $\sim120$ mm.  Because geometric angles are magnified with respect to on-sky angles, this relatively large shift in height corresponds to an elongation of 0.6 Hartmann pixels (or one spot size of $2\farcs0$).  A full-aperture image of the elongated spots is shown in Figure 5; notice that the spots are defocused at extrema, creating a characteristic ``peanut'' shape for side guide stars.  The principal effect of elongation in on-sky systems is to reduce centroiding accuracy and increase measurement error.  However, because this bench uses bright guide stars, the measurement error is largely not affected.  The elongation actually decreases WFS systematic error due to WFS nonlinearity effects \citep{amm07}.  As the effect of elongation is slight, we do not perform this step for the experiments presented in this paper.
\begin{figure}
   \begin{center}
   \begin{tabular}{c}
   \includegraphics[height=6cm]{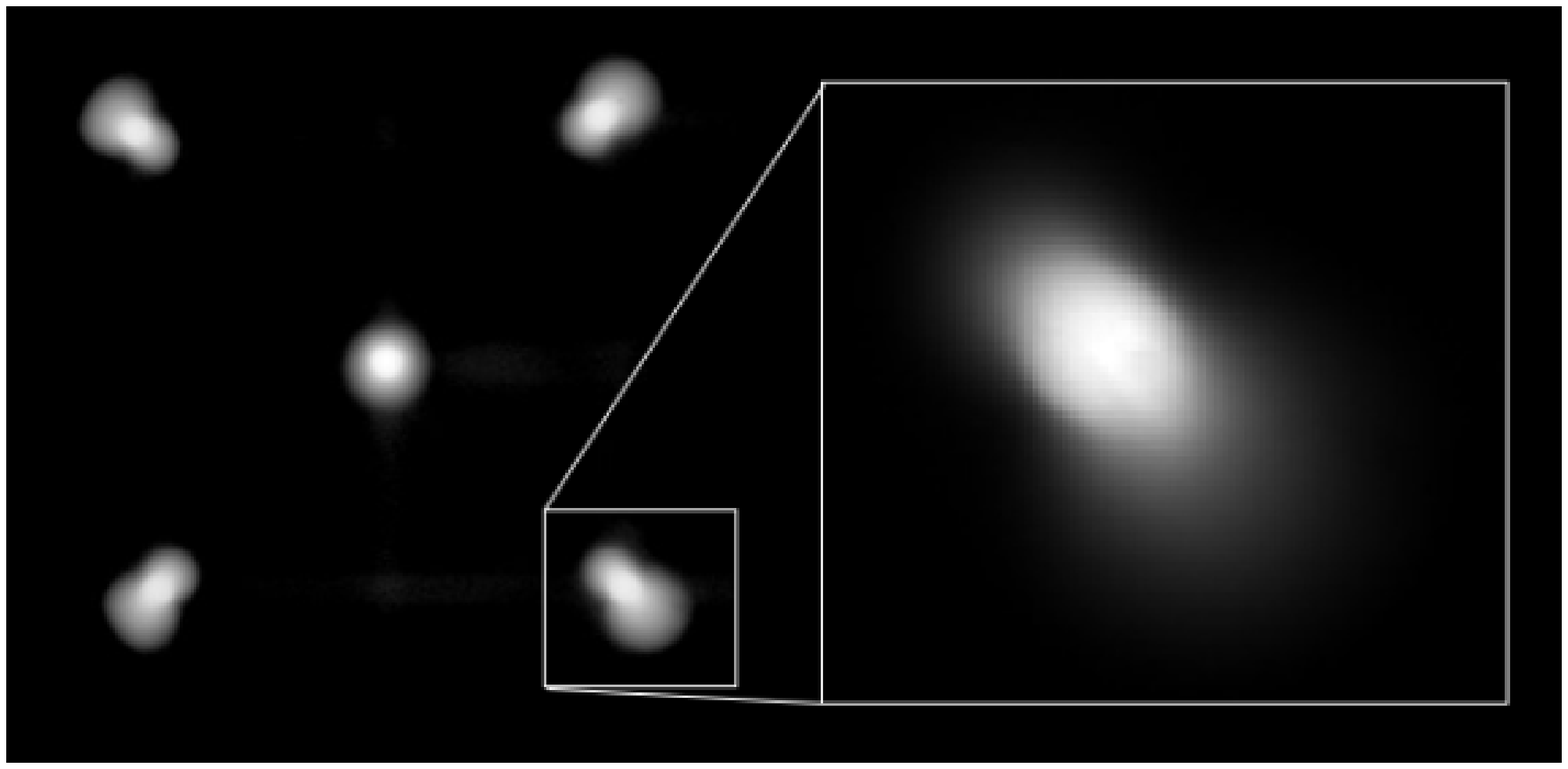}
   \end{tabular}
   \end{center}
   \caption[example] 
{ \label{fig:elongation_image} 
Full-aperture images of laser guide stars during mechanical elongation.  The left panel is in log scale and the inset is in linear scale.  The principal effect of axial translation is defocusing rather than lateral motion for side guide stars (compared to on-sky LGS systems), as the geometric angles in this testbed are magnified compared to the angular dimensions of a 10-meter telescope.  This creates a characteristic ``peanut'' shape for side guide stars.}
   \end{figure}

\subsection{Atmospheric Phase Plates}

We use four different types of phase plates:  Etched glass, acrylic-coated glass, CD cover, and hairspray-coated glass.  All types have Kolmogorov power spectra extending to the spatial resolution limit of our test interferometer (5 pixels per Hartmann subaperture).   The CD covers tend to have more non-Kolmogorov structure, including excesses of astigmatism and coma, but they have the advantage of reaching higher total RMS variation in a 2-cm metapupil (0.5 - 1.0 microns RMS) than the other types.  They also display less amplitude variation than the other types.  The etched glass plates were manufactured at LLNL and have pure Kolmogorov power spectra etched into them.  However, these plates possess an excess of high spatial frequencies that introduce WFS measurement error.   For the weaker plates, this error is small, as is the amplitude variation across the plate, so we use these weaker plates to simulate high altitude layers.   We distribute the atmosphere in three layers (0, 4.5, and 9 km) with strengths that mimic Mauna Kea-type $C_n^2$ atmospheric profiles (the exact distribution is presented in Section 3).

The acrylic- and hairspray-coated plates were manufactured in the Laboratory for Adaptive Optics at UCSC.  These plates largely possess Kolmogorov power spectra, but ``bubble''-like features are frequent.  These features are manifest as regions of low intensity in the WFS sensors, which introduces measurement error.   These non-Kolmogorov bubbles are easily picked out and avoided. 

The optimum plates for the ground layer are CD covers, as they are the only type with sufficient phase amplitude to model dominant layers.  For layers at high equivalent altitude, it is important that total RMS phase error be low ($< 150$ nm) to reduce scintillation error terms.  Total intensity variation must be low for all layers, as the Hartmann sensor is sensitive to intensity dropouts.   Power spectra of phase for the various plates used are shown in previous publications \citep{amm07}.

\subsection{Insertable Field Stops}

The high-order deformable mirrors (Programmable Phase Modulators, or PPMs) produce a grid of diffraction spots due to their pixelated nature.  The intensity in these diffraction peaks varies with the particular PPM used and its displayed phase shape.  Unfortunately, the physical separation between the zeroth order and the first order is equivalent to $45\farcs0$, approximately the same as the laser guide star separation.  We rotate the LGS constellation 22 degrees from the proper ``quincunx'' pattern to minimize the interference between the LGS diffraction spots and the PPM rotation axis.   During the wavefront sensing step, we insert field stops with five holes of 1.6 mm diameter to pass the 5 laser guide stars.  These holes are several times larger than the spatial filters used to improve wavefront sensing in other systems.  For example, the Gemini Planet Imager \citep{mac06} uses small spatial filters to eliminate high spatial frequencies that contribute to aliasing errors.  

\subsection{Deformable Mirrors and Phase-Wrapping}

Three deformable mirrors are present in this system:  A ground layer woofer (ALPAO DM-52), a Hamamatsu Programmable Phase Modulator (PPM) at the ground layer, and another PPM device at an equivalent altitude of 9 km.  The PPMs have very high spatial resolution, with 768x768 independent controllable pixels across the screen.  An approximately 5x5 region on the PPM maps to one Shack-Hartmann subaperture.  The ground-conjugated woofer-tweeter pair and the second PPM at higher altitude can be used for closed-loop MCAO experiments, the results of which have been previously published \citep{laa08}.  

We now use the system for MOAO experiments by flattening the ground-layer PPM and woofer during wavefront sensing, so as to sense the atmosphere wholly in open-loop.  Both the PPM and woofer are highly repeatable on short timescales, so very little error is introduced by performing MOAO in this manner ($< 5$ nm RMS).   For MOAO, the 9 km PPM is powered down at all times.  The ground-layer PPM is the only deformable mirror that corrects introduced atmospheric turbulence; the woofer is only used to correct instrumental errors, as described below.  

The influence functions of the PPM pixels are much smaller in width than our Hartmann subapertures, so the transformation of desired phase into pixel commands is more direct than with traditional deformable mirrors.  The desired phase is simply interpolated onto the PPM pixel map.  The open-loop response of the PPMs to voltage signals has been calibrated across its full range with a Quadrature Polarization Interferometer (QPI).  These signals have been stored for each pixel, inverted, interpolated, and used to form a look-up table.  The QPI measurement noise of $\sim30$ nm RMS propagates into the PPM lookup tables; thus the PPMs can be used in open-loop with a go-to error of $\sim30$ nm.

The PPMs also have a limited stroke of $600-900$ nm.  In order to create shapes with typical atmospheric amplitudes, we wrap the phase of the PPMs, typically 8-15 times in a pupil.  This phase-wrapping is essentially the same as performing a modulus operation on the desired phase map.  This is optically permissible for two reasons:  (1) The spectral width of the laser guide stars and science test stars is negligibly small, so the $2\pi$ phase jumps appear to be smooth transitions; and (2) the phase transitions are sharp because the PPM influence functions are narrow.  As a result, this phase-wrapping does not affect point spread function (PSF) formation.  

The woofer mirror is used primarily in these experiments to introduce a field-dependent static correction, which does not change during the course of experimentation.  It is not used to assist the ground-layer PPM in correcting the atmospheric turbulence.  Each field point has a separate static correction wavefront, which is only placed on the woofer when Strehl is checked on that star.  The 9 km PPM is not used for static correction, but is left unpowered.  During wavefront sensing, both the woofer and ground-layer PPM are flattened.  

We store a field-dependent ``static correction'', or low-order pattern, that cancels the astigmatism introduced by a 3 inch beamsplitter and other misaligned optics within the system.  The static correction is generated by a combination of two techniques:  Automated image sharpening of low-order modes, up to fourth Zernike order, and image sharpening via dithering of individual woofer actuators.  The modal sharpener fits the science PSF to a perfect Airy model to estimate the merit of correction.  The dithering sharpener attempts to maximize the Strehl itself.  We also use the Quadrature Polarization Interferometer (QPI), coupled to the MOAO/LTAO system, to measure and cancel high-order aberration caused by the PPMs themselves.   The combination of these methods gives Strehls of $83\%$ across the science field, when optimized for individual stars.  

\subsection{Software and Control}

The control algorithms for this testbench are discussed in more detail in prior publications \citep{amm06, amm07, laa08} and here summarized.  No part of the reconstruction or control is matrix-based.  We use simple center-of-mass centroiding over 4x4 pixel subapertures, followed by a slope linearization lookup (see section 2.3).  The slopes are reconstructed into phase using Fourier techniques \citep{poy03}.   This reconstruction technique is performed iteratively to reduce edge effects.  Mean piston and tip/tilt are then removed from the five measured pupils.  The edge phase values are ``smeared'' several subapertures beyond the edge of the pupil to further reduce edge errors.  The pupils are dewarped and registered before being analyzed tomographically.  The tomographic analysis code is ``Tomography Spherical Wave,'' (TSW) which applies a minimum variance preconditioned conjugate gradient (PCG) solver to the problem of back-projection tomography \citep{gav04}.  We use alternating sets of 5 PCG and Fixed iterations, for a total of 35 iterations, to provide the tightest convergence.  The number and strength of ``software'' atmospheric layers in TSW correctly models the simulated atmosphere for all realizations.  

Following the tomographic procedure, the optimized phase corrections are placed on the DM.  For each science star, a different shape is placed on the DM to correct turbulence and field-dependent static errors.  After checking Strehl on all stars independently, the wind motors are driven and the next AO iteration begins.  

\subsection{Similarity Parameters}

\subsubsection{Geometric Scaling} 

The problem of constructing a testbed that simulates execution of wide-field adaptive optics on a 10-meter telescope begins with several geometric constraints.  High-order deformable mirrors with good go-to performance are commercially available with sizes of order 1-2 cm, restricting the simulated telescope aperture to nearly this size when large magnification factors are avoided.  In addition, atmospheric turbulence plates are readily available with Fried parameters of $200-500\; \mu$m at optical wavelengths, permitting $D/r_0$ scales of $30-60$ with a $1-2$ cm pupil, suitable for investigating visible light AO.   However, the simulated atmospheric space must represent $\sim15$ equivalent kilometers to fully simulate three-dimensional atmospheric profiles.  Matching the transverse/longitudinal aspect ratio to that of a real 10-meter telescope would require this space to be on the order of 15 meters, occupying lab space unnecessarily.  The MOAO/LTAO testbed compresses the height dimension in the atmosphere by a factor of 60 to address this.  The height scale, LGS constellation mount, and atmospheric plate locations are set to mimic the beam footprints of a true LGS constellation at the relevant atmospheric heights.  

Scintillation error effects, due to Fresnel propagation of phase errors at different atmospheric heights into amplitude variation, are worsened for scaled-down testbeds.  This is because the ratio of the size of the entrance pupil to the wavelength of light is much smaller than on real telescopes.  As seen in the scintillation index equation in section 4.3.2, further compressing of height scales mitigates this effect.  

For simulation of wide-field AO, it is most important that the following geometric characteristics of the system are analogous to those of a real telescope:  (1) Wavefront sensor geometry and subaperture number, (2) atmospheric strength and distribution across equivalent atmospheric heights, (3) deformable mirror actuator size and DM-lenslet mapping, (4) LGS and science star metapupils for all atmospheric layers, and (5) LGS spot size and range of motion as seen by Shack-Hartmann subperatures.  For the MOAO/LTAO system, these characteristics match a hypothetical wide-field AO system on a 10-meter telescope with circular aperture, except that the LGS size is 0.61 pixels rather than $\sim1$ pixel, which only affects Hartmann linearity.  

The critical system parameters, in both physical laboratory units and on-sky equivalent units, are presented in Table 1. 

\begin{table}[h]
\caption{Critical mapping parameters for testbed and corresponding on-sky simulation.} 
\label{tab:simparams}
\begin{center}       
\begin{tabular}{|l|l|l|} %% this creates two columns
%% |l|l| to left justify each column entry
%% |c|c| to center each column entry
%% use of \rule[]{}{} below opens up each row
\hline
\rule[-1ex]{0pt}{3.5ex}  Parameter & Atmosphere & Lab \\
\hline
\hline
\rule[-1ex]{0pt}{3.5ex}  Transverse magnification & 1 & 0.00183 \\
\hline
\rule[-1ex]{0pt}{3.5ex}  Longitudinal magnification & 1 & $1.67\times10^{-5}$ \\
\hline
\rule[-1ex]{0pt}{3.5ex}  Atmospheric path length & 9 km & 150 mm \\
\hline
\rule[-1ex]{0pt}{3.5ex}  Aperture, D & 10 m & 18.3 mm \\
\hline
\rule[-1ex]{0pt}{3.5ex}  Field Diameter & $42\farcs5$ & $42\farcm5$  \\
\hline
\rule[-1ex]{0pt}{3.5ex}  Subaperture size, d & 15.2 cm & 278 $\mu$m \\
\hline
\rule[-1ex]{0pt}{3.5ex}  Mean $r_0$ at 0.5 $\mu m$ & 16.2 cm & 296 $\mu$m \\
\hline 
\rule[-1ex]{0pt}{3.5ex}  Mean $\tau_0$ at 0.5 $\mu m$ & 3\farcs17 & 3\farcm17 \\
\hline
\rule[-1ex]{0pt}{3.5ex}  Science wavelength & 0.71 $\mu$m & 0.658 $\mu$m \\
\hline
\rule[-1ex]{0pt}{3.5ex}  $D/r_0$ at science $\lambda$ & 41.41 & 41.41 \\
\hline 
\rule[-1ex]{0pt}{3.5ex}  Mean RMS Atmospheric strength & 921 nm & 854 nm \\
\hline
\end{tabular}
\end{center}
\end{table}  

\subsubsection{Wavelength Scaling} We quantify AO performance at the laser wavelength used for the science stars, $658$ nm.  The performance (i.e., Strehl) of the AO system is largely set by the atmospheric strength, specified by $D/r_0$, with $r_0$ given at the science wavelength.  In turn, $r_0$ is a function of the simulated seeing strength ($r_0$ at 500 nm, or $r_{0,500}$) and the simulated science wavelength ($\lambda$).  We simulate the AO performance at science wavelengths other than 658 nm by choosing a $D/r_0$ that corresponds to median seeing at a different wavelength.   The following equation models the dependency of total atmospheric strength on $D/r_0$ \citep{har98}:
$$\sigma = 0.366\;( D/ r_0 )^{5/6},$$
where $\sigma$ is the tip/tilt-removed RMS deviation of phase in units of laboratory radians (i.e., $658/2\pi = 105$ nm) and $r_0$ is the Fried parameter at the science wavelength.  $r_0$ is dependent on the seeing strength in the following way \citep{har98}:
$$r_0 = r_{0,500} \;(\lambda / 500\; nm)^{6/5}.$$
Then
$$\sigma = 0.366\; (500\;nm / \lambda) \;( D / r_{0,500})^{5/6}.$$
With a 10 meter telescope, a $\sigma$ of 9 radians (950 nm) RMS corresponds to $r_0 = 21.4$ cm, simulating median seeing at optical red wavelengths ($r_{0, 500} = 15.4$ cm for $\lambda = 658$ nm) or poor seeing at infrared wavelengths ($r_{0, 500} = 9.3$ cm for $\lambda = 1.0\;\mu m$).  There are some inaccuracies inherent to this model.  Most of the crucial error budget terms scale linearly with wavelength, but several do not, including the static uncorrectable errors ($\sim45$ nm for the system).  As a result, simulations at wavelengths bluer than $658$ nm overpredict the Strehl and simulations at wavelengths redder than $658$ nm underpredict.   Because of these inaccuracies, we choose a science wavelength as close to $658$ nm as possible while still preserving system performance.  Setting the total amplitude of phase error to $\sim850$ nm RMS in turn sets the science wavelength to 710 nm for median seeing ($r_{0, 500 nm} = 16.2$ cm), permitting R-band tests.  

\subsubsection{Temporal Scaling}

All physical processes that govern the AO performance, if they are controllable, can be slowed (or stopped and restarted) during operation without affecting performance.  Real computation time can be freely extended without degrading performance as long as the atmosphere stays frozen during operation.  In addition, we eliminate the \textit{delay} errors present in on-sky AO systems by keeping the atmosphere frozen during the wavefront sensing, computation, and deformable mirror actuation steps.  Each AO iteration uses 8 seconds in real time to measure wavefront sensor signals, perform reconstruction and tomographic analysis, and apply deformable mirror shapes.  Measuring Strehl uses an additional 40 seconds, as different DM shapes must be applied and new science images taken for each of seven science stars.  The length of equivalent time that each AO iteration simulates is set \textit{only} by the physical speeds of the atmospheric plates (in microns per iteration) and the desired simulated wind speed (in meters per second).  As discussed below, we choose plate speeds that simulate AO operation at 1 kHz.

\section{PERFORMANCE AT VISIBLE WAVELENGTHS}

\subsection{Experimental Makeup}
Previous work with this testbed has demonstrated that open-loop operation decouples instantaneous AO performance from preceding AO iterations \citep{amm08}.  The only dependence of performance on simulation timepoint arises from a gradual degradation of Hartmann references and open-loop wavefront sensor calibration accuracy with time, caused by internal air movement and optical drift induced by slow temperature relaxation.  These dependencies introduce small error terms, which we explicitly model in Section 4.3.

In this section, we present experimental results from four individual manifestations of the atmosphere, with no continuous transitional evolution between the four states.   This approach maximizes the atmospheric variation in the experiment.  For each realization, we evolve the atmosphere with a variable Taylor frozen-flow model over 200 equivalent milliseconds.  Although the wind speeds are set to simulate operation at 1 kHz, the instantaneous performance is independent of the Hartmann frame rate, unlike real AO systems on telescopes.   This is because the bandwidth and delay error terms are not included in this simulation.  We perform AO iterations every 5 milliseconds, giving a record of atmospheric statistics and the performance of the tomographic algorithm at a 200 Hz rate; we check the Strehl on the science stars every 10 milliseconds to give AO performance data at a 100 Hz sampling rate.   The trends in AO performance measured would be unchanged, only more densely sampled, if we performed AO iterations every 1 millisecond.  

Overall, this amounts to nearly one full second of simulation data ($800$ ms) comprising four atmospheric realizations, $160$ AO iterations, and $80$ Strehl measurements.  This models the common practice of recording on-sky telemetry on random nights throughout a year of operation, as the phase in different atmospheric realizations is distinct.

\subsection{System Performance}

The atmospheric parameters and performance for the four realizations, including R-band Strehls, ensquared energies, and wind speeds, are listed in Table 2.  Over the course of these experiments, the system attains mean R-band Strehls of $32.4\%$ on-axis, $24.5\%$ at a distance $10\farcs0$ off-axis, and $22.6\%$ at a distance of $15\farcs0$ off-axis.  The mean ensquared energy for 50 milliarcsecond spaxels on-axis is 46\%.  The distributions of $r_0, \tau_0,$ and Strehls for all stars are shown in Figure 6.  Notice that the Strehls quickly rise to equilibrium values after the start of the simulation; this is a natural feature of open-loop AO operation.  

\begin{figure}
   \begin{center}
   \begin{tabular}{c}
   \includegraphics[height=7.5cm]{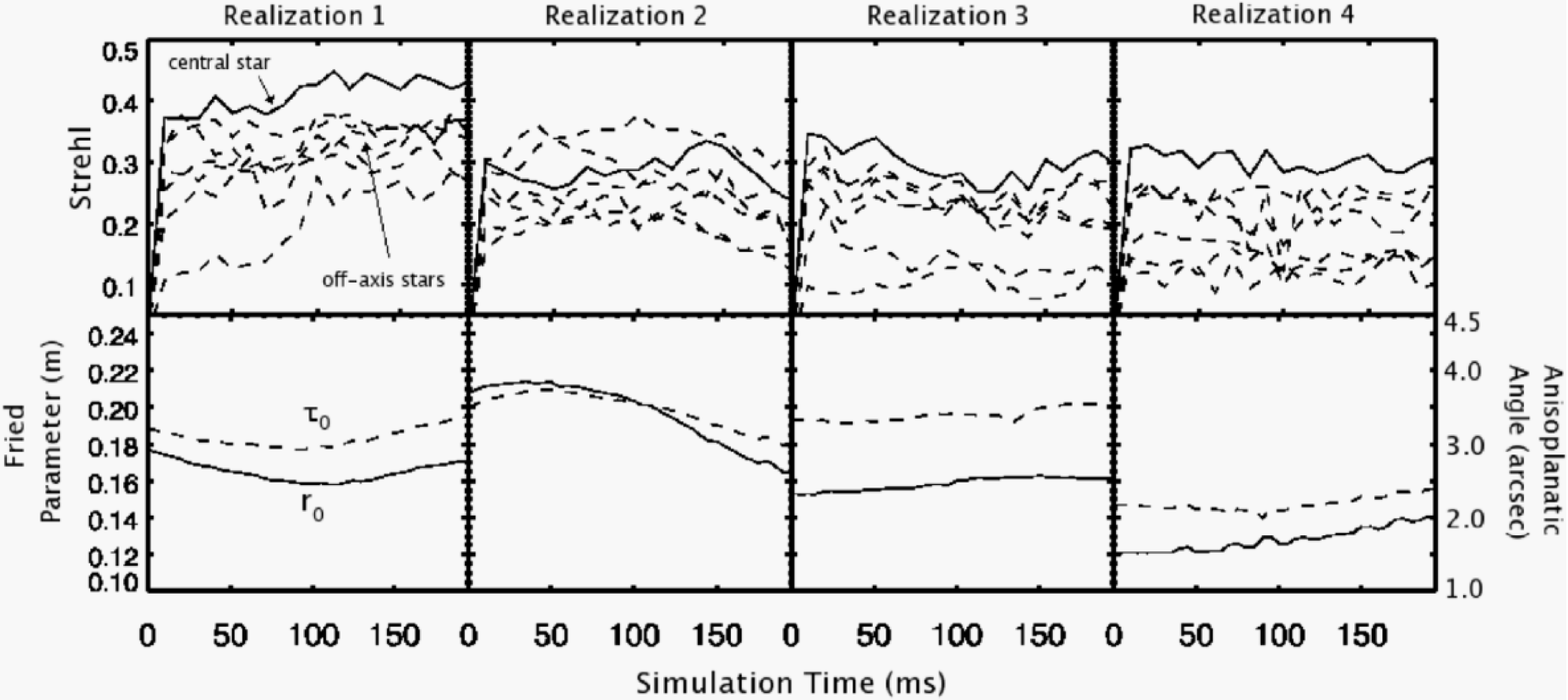}
   \end{tabular}
   \end{center}
   \caption[example] 
{ \label{fig:strehl_plot} 
R-band Strehls and atmospheric parameters versus simulation time.  Strehls are shown in the upper four panels and $r_0$ and $\tau_0$ are plotted in the lower panels.  The y-axes units for $r_0$ are shown on the left side of the plot and the units for $\tau_0$ are shown on the right side.  Data for the four atmospheric realizations are presented left-to-right.  The solid lines in the top plots denote on-axis strehls; dashed lines denote off-axis strehls ($10\farcs0 - 15\farcs0$).  The solid lines in the lower panels plot the Fried parameter $r_0$ and the dashed lines plot the anisoplanatic angle $\tau_0.$}
   \end{figure} 
   
Time-averaged point spread functions are shown for three off-axis distances in Figure 7.  The image FWHM's range between $0\farcs02$ (the diffraction-limit at this wavelength) and $0\farcs03$.  Using the seven science test stars, it is possible to map the nominal Strehl as a function of position in the MOAO field and compare it to a normal single-conjugate AO system at this wavelength.   Figure 8 displays MOAO and anisoplanatic Strehl maps.  The MOAO Strehl map is obtained by interpolating through the mean Strehls obtained in the experiments discussed above.  The map of anisoplanatic Strehls is obtained in the following manner:  A 3D atmosphere is set up, as for MOAO, and a full tomographic wavefront sensing analysis is performed.  Line integrals are performed through the reconstructed atmosphere for a large number of field points surrounding the central science star.  When these integrated wavefronts are placed on the ground-layer deformable mirror, the Strehl on the central star is checked and mapped as a function of the field point.  This approximates the anisoplanatism at the chosen science wavelength (710 nm).  The Strehl distribution (shown in the right panel of Figure 8) is not radially symmetric because it is not time-averaged over multiple atmospheric realizations, but it is a useful indication of the magnitude of anisoplanatism.  If an R-band Strehl cutoff of $15\%$ is chosen to signify ``usable'' imagery, then the MOAO field of regard has been expanded to $0.5$ square arcminutes, or $\sim25$ times larger than that limited by anisoplanatism.

\begin{figure}
   \begin{center}
   \begin{tabular}{c}
   \includegraphics[height=8cm]{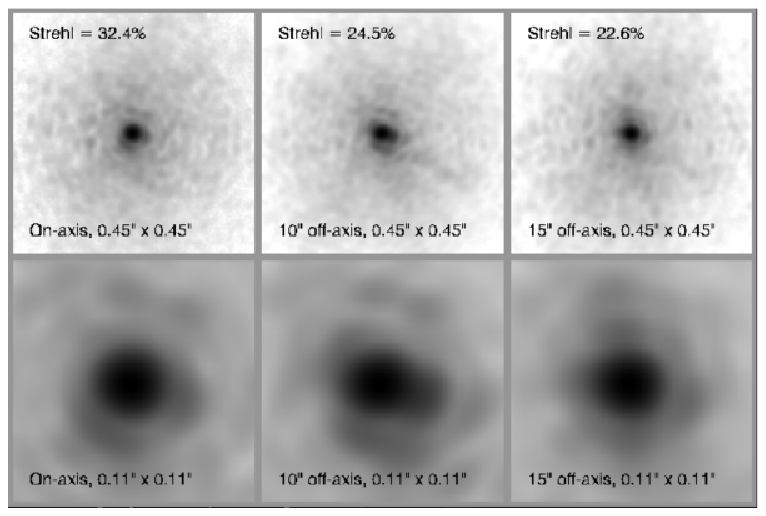}
   \end{tabular}
   \end{center}
   \caption[example] 
{ \label{fig:all_psfs} 
Inverted logarithmic images of mean point spread functions at three off-axis distances.  The left panels show the PSF on-axis, the middle panels show the PSFs at $10\farcs0$ off-axis, and the right panels display the PSFs at $15\farcs0$ off-axis.  The top panels have image sizes of $0\farcs45 \times 0\farcs45$.  The bottom panels have image sizes of $0\farcs11 \times 0\farcs11$.  R-band Strehls are displayed in the top panels.}
   \end{figure} 

\begin{figure}
   \begin{center}
   \begin{tabular}{c}
   \includegraphics[height=7cm]{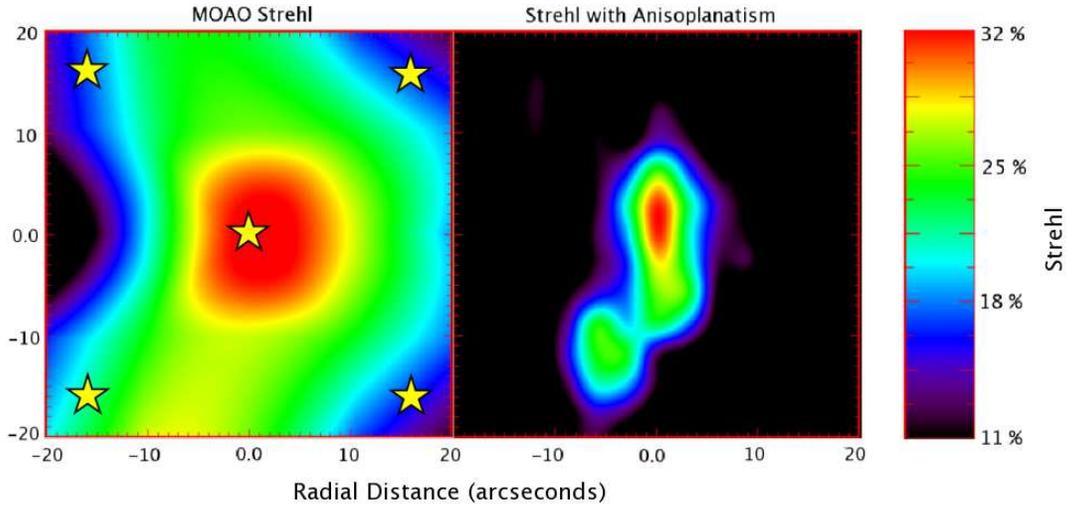}
   \end{tabular}
   \end{center}
   \caption[example] 
{ \label{fig:strehl_map} 
MOAO R-band Strehl map and anisoplanatic Strehl map.  The MOAO R-band Strehl map is an interpolation to the mean results from the four atmospheric realizations presented here.  The yellow stars indicate the locations of laser guide stars.  The anisoplanatic Strehl map is obtained as described in the text for only one atmospheric realization.}  
   \end{figure} 

 \begin{table}[h]
\caption{System performance for 10 meter case.  Strehls, wind speeds, and layer strengths are given for four atmospheric realizations column-wise.  The ``Mean EE (50 mas)'' values denote ensquared energies for 50 milliarcsecond spaxel sizes.} 
\label{tab:performance}
\begin{center}       
\begin{tabular}{|l|l|l|l|l|} %% this creates two columns
%% |l|l| to left justify each column entry
%% |c|c| to center each column entry
%% use of \rule[]{}{} below opens up each row
\hline
\rule[-1ex]{0pt}{3.5ex}  Realization & 1 & 2 & 3 & 4\\
\hline
\hline
\rule[-1ex]{0pt}{3.5ex}  $r_0$ (cm) & 16.5 & 19.7 & 15.8 & 12.8 \\
\hline
\rule[-1ex]{0pt}{3.5ex}  $\tau_0$ ('') & 3.23 & 3.63 & 3.55 & 2.27 \\
\hline
\rule[-1ex]{0pt}{3.5ex}  Mean Strehl on-axis & 41.0\% & 28.6\% & 29.8\%&30.2\% \\
\hline
\rule[-1ex]{0pt}{3.5ex}  Mean Strehl 10'' off-axis& 32.3\% & 21.9\% & 24.4\%&19.5\% \\
\hline
\rule[-1ex]{0pt}{3.5ex}  Mean Strehl 15'' off-axis & 29.1\% & 25.8\% & 17.8\%&17.8\% \\
\hline 
\rule[-1ex]{0pt}{3.5ex}  Mean EE (50 mas) on-axis & 51.3\% & 47.0\% & 44.2\%&41.5\% \\
\hline 
\rule[-1ex]{0pt}{3.5ex}  Mean 0 km strength & 49.3\% & 41.0\% & 68.3\%&54.2\% \\
\hline 
\rule[-1ex]{0pt}{3.5ex}  Mean 4.5 km  strength & 37.3\% & 46.8\% & 16.3\%&31.7\% \\
\hline 
\rule[-1ex]{0pt}{3.5ex}  Mean 9 km strength & 13.4\% & 12.2\% & 15.4\%&14.1\% \\
\hline 
\rule[-1ex]{0pt}{3.5ex}  0 km wind speed (m/s)& +6.1 & +7.9 & +4.8 & +3.0\\
\hline
\rule[-1ex]{0pt}{3.5ex}  4.5 km wind speed (m/s)& -15.2 & -17.0 & +9.1 & -7.0\\
\hline
\rule[-1ex]{0pt}{3.5ex}  9 km wind speed (m/s)& +20.0 & -12.1 & -15.2 & -7.9\\
\hline
\rule[-1ex]{0pt}{3.5ex}  Simulation length (ms) & 200 & 200 & 200 &200 \\
\hline

\end{tabular}
\end{center}
\end{table} 

\section{ERROR BUDGET}

\subsection{Summary}
In this section we present a time-dependent error budget, comprising 13 separate static and parameterized error models, which we use to predict the spatial and temporal performance of the AO system.  Each of the error terms have been verified with independent, empirical measurements, or justified with separate numerical simulations.  The error terms and models are discussed in section 4.3.  The full error budget is presented in Table 3, including on-axis cases in both laboratory and on-sky units and an off-axis case in on-sky units.

This testbed studies the idealized performance of an MOAO/LTAO system, with bright tip/tilt loops and no delay or bandwith error.  However, it is subject to a large number of static and calibration errors discovered in existing AO systems, including:
\begin{enumerate}
\item[(1)]Static and dynamic uncorrectable modes  (largely introduced by the Programmable Phase Modulator deformable mirrors, or PPMs)
\item[(2)]Static and dynamic WFS zero-point calibration error
\item[(3)]Go-to control errors (the open-loop control error measured for the PPM is set by the measurement accuracy of the quadrature polarization interferometer, used to calibrate it)
\item[(4)]A particular manifestation of DM finite stroke errors (errors due to insufficiency of PPM wrapping to completely describe a unwrapped wavefront, due to the spatial smoothing of the phase on the PPM)
\item[(5)]DM-to-lenslet warping and misregistration error
\end{enumerate}

 \begin{table}[h]
\caption{Error budget for 10 meter case at $710$ nm.  Each term represents the mean over all $160$ AO iterations over 4 atmospheric realizations.  All values are in units of nanometers, unless otherwise stated.  The left column gives error budget values in on-sky units on-axis.  The middle column gives values in on-sky units at an off-axis distance of $12\farcs5$.  The right column displays error budget values in physical, laboratory units (unstretched) for the on-axis case.} 
\label{tab:budget}
\begin{center}       
\begin{tabular}{|l|l|l|l|} %% this creates two columns
%% |l|l| to left justify each column entry
%% |c|c| to center each column entry
%% use of \rule[]{}{} below opens up each row
\hline
\rule[-1ex]{0pt}{3.5ex}  Error Budget Term & On-sky, on-axis & On-sky, off-axis ($12\farcs5$) & Lab, on-axis (658 nm) \\
\hline
\hline
\rule[-1ex]{0pt}{3.5ex}  Fitting Error & 40.7 & 40.7 & 37.7 \\
\hline
\rule[-1ex]{0pt}{3.5ex}  WFS Aliasing & 16.2 & 16.2 & 15.0 \\
\hline
\rule[-1ex]{0pt}{3.5ex}  Tomography Error & 69.0 & 83.8 & 63.9 \\
\hline
\rule[-1ex]{0pt}{3.5ex}  WFS Systematic error & 41.5 & 41.5 & 38.5 \\
\hline
\rule[-1ex]{0pt}{3.5ex}  Field Stop Misalignment & 10.8 & 10.8 & 10.0 \\
\hline 
\rule[-1ex]{0pt}{3.5ex}  PPM Lookup Table error & 32.4 & 30.0 & 30.0 \\
\hline 
\rule[-1ex]{0pt}{3.5ex}  Static Uncorrectable, S=83\%& 48.6 & 45.0 & 45.0 \\
\hline 
\rule[-1ex]{0pt}{3.5ex}  Scintillation & 12.6 & 12.6 & 11.7 \\
\hline 
\rule[-1ex]{0pt}{3.5ex}  WFS Scintillation & 26.8 & 26.8 & 24.8 \\
\hline
\rule[-1ex]{0pt}{3.5ex}  Photon error & 16.2 & 16.2 & 15.0 \\
\hline
\rule[-1ex]{0pt}{3.5ex}  WFS zeropoint drift & 10.8 & 10.8 & 10.0 \\
\hline
\rule[-1ex]{0pt}{3.5ex}  Linearity calibration drift & 10.8 & 10.8 & 10.0 \\
\hline
\rule[-1ex]{0pt}{3.5ex}  Pupil registration drift & 25.9 & 25.9 & 24.8 \\
\hline
\hline
\rule[-1ex]{0pt}{3.5ex}  Total RMS & 118.2 & 127.3 & 109.5 \\
\hline 
\rule[-1ex]{0pt}{3.5ex}  Predicted Strehl (\%) &  33.5 & 28.1 & 33.5 \\
\hline 
\rule[-1ex]{0pt}{3.5ex}  Measured Strehl (\%) & 32.4 & 23.6 & 32.4 \\
\hline 
\rule[-1ex]{0pt}{3.5ex}  Relative Error in Model & 3.3\% & 16.0\% & 3.3\% \\
\hline
\end{tabular}
\end{center}
\end{table} 

Using the known atmospheric strengths at every timepoint in the MOAO experiments, it is possible to predict the Strehl as a function of time and space from this error budget.  These predictions are shown in Figure 9 for the four realizations.   These Strehl plots should be compared against the actual Strehls measured.  Figure 10 compares the predicted Strehls against the measured Strehls as a function of atmospheric realization and field angle, integrated over simulated time.  Averaged over all realizations, the on-axis predictions match the measured Strehls to 3\%, but our error budget model overpredicts the off-axis Strehl by $\sim16\%$.   This is likely due to propagation of WFS systematic errors through the tomographic algorithm, which become manifest as ``rings'' of error at the edges of the upper altitude metapupils.  This propagation was not modeled in the error budget.

\begin{figure}
   \begin{center}
   \begin{tabular}{c}
   \includegraphics[height=8.5cm]{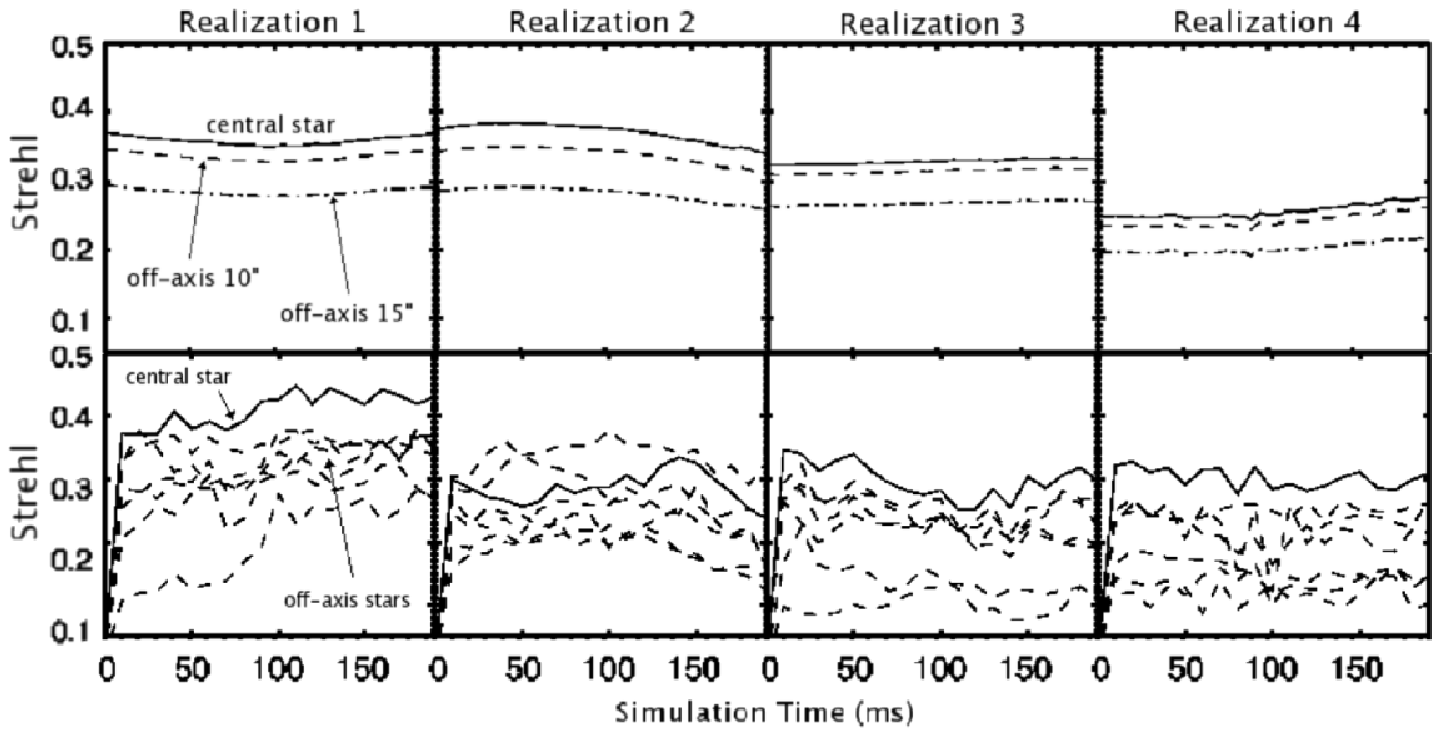}
   \end{tabular}
   \end{center}
   \caption[example] 
{ \label{fig:error_predict} 
Top series:  Strehl predictions based on a complete time- and space-dependent error model.  Bottom series:   Actual Strehl measurements, as replicated from the top half of Figure 6.  In all panels, solid lines denote the predicted AO performance on-axis.  In the top series, dashed lines show the predicted performance at a field radius of $10\farcs$  Dot-dashed lines plot the predicted Strehls at a radius of $15\farcs$  The two off-axis predictions should be compared against the dashed lines in the bottom panels, denoting the measured off-axis Strehls.}  
   \end{figure} 

\begin{figure}
   \begin{center}
   \begin{tabular}{c}
   \includegraphics[height=5.2cm]{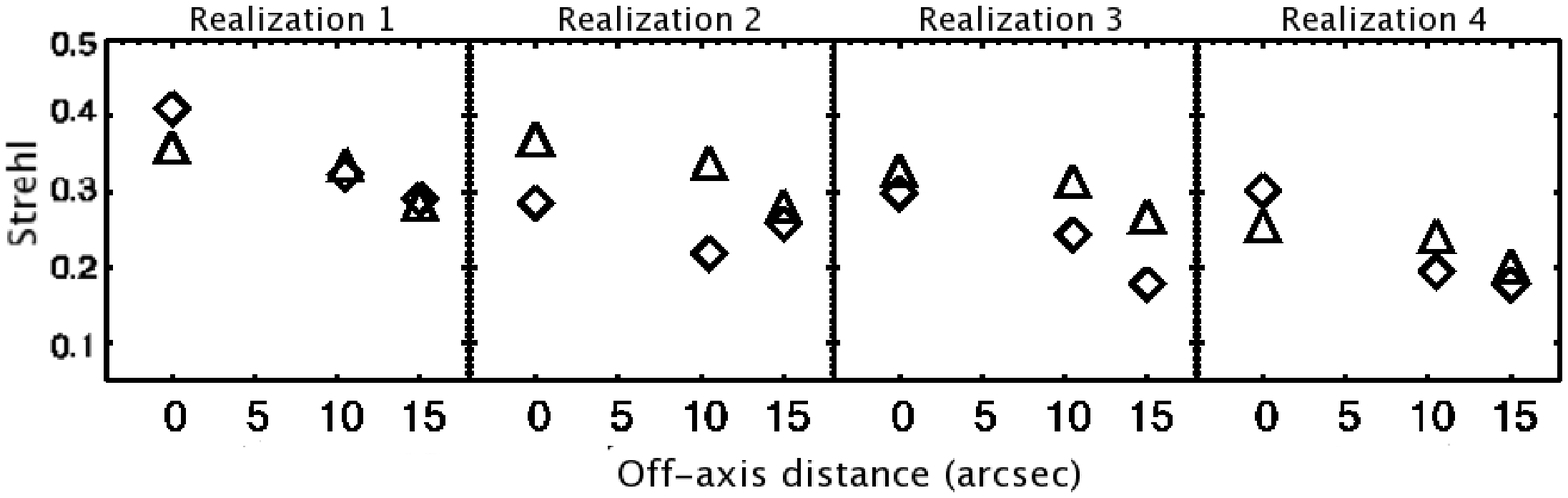}
   \end{tabular}
   \end{center}
   \caption[example] 
{ \label{fig:strehl_predict} 
Strehl predictions plotted against measured Strehls.  Panels are arranged by atmospheric realization horizontally.  Plotted on the x-axis is the field position in arcseconds and plotted on the y-axis is the Strehl.   The predicted values are shown as triangles, and the measured values are shown as diamonds.  Note that the on-axis predictions match the measured values, but the off-axis predictions are $\sim16\%$ higher than the measured values. }  
   \end{figure} 

\subsection{Errors Not Modeled}

This testbench was built to investigate the idealized performance of Laser Tomographic Adaptive Optics (LTAO) and similar technologies.  The photon error in the system ($\sim15$ nm RMS per pupil) models 30W sodium-layer laser guide stars sharpened to $0\farcs8$ and a low read-noise CCD.  We are not subject to bandwidth or delay error because (1) the system does not pass or record science light between AO iterations and (2) there is no time delay between wavefront measurement and correction.

Tip/tilt errors cannot be measured by laser guide stars because the emitted photons re-trace their own upward path through the atmosphere.  Multi-LGS systems require multiple natural guide stars to constrain global tip/tilt as well as quadratic modes unsensed by the laser guide stars.  Because we are free to take individual science frames every AO iteration, we time-average the science PSFs by shifting-and-adding individual images to remove tip/tilt errors explicitly.  This models a case with a bright tip/tilt star, for which there is little residual tip/tilt jitter.  All Strehls quoted in this paper are measured from instantaneous (not time-integrated) PSFs, although the time-averaged PSFs are displayed in Figure 6.  The shifting-and-adding sufficiently models the global tip/tilt loop in the bright star case, but the testbed is still subject to coupling of low-order, high altitude quadratic modes into guide star tip/tilt.  This introduces an error term in the overall error budget, which is included in the ``tomographic error'' term.  It is envisioned that tip/tilt measurements from multiple natural guide stars will address this error term, as modeled in the split tomography algorithm of \citet{gil08}, for example.

\subsection{Individual Error Models}

We now present the details of the time- and space-dependent error model used to predict the Strehls in Figure 10.  The mean values of each of these terms are shown in Table 3.   We stress that the majority of these terms have been measured with independent, empirical methods.  We now group these error terms by their dependence on atmospheric strength (and therefore time) or field radius.

\subsubsection{Static Terms}
Nearly half of the error budget terms are static, meaning that they have no time- or space- dependence.  The first of these is \textit{measurement error} due to photon counting noise.  The static value of $\sim15$ nm is determined by measuring the RMS disagreement between measured wavefronts from iteration to iteration.  This value is quite low compared to current on-sky systems, but similar to that expected in the realistic case of bright laser guide stars ($\sim30$ W) sharpened by low-order uplink correction to the Hartmann pixel size ($0\farcs8$).  

The second static term is the \textit{PPM lookup table} error.  This error reflects the measurement error of the QPI, used to characterize the PPM deformable mirrors.  Because this error is manifest over one full wave of range, and we wrap phase by $2\pi$ to achieve greater stroke, it is not proportional to atmospheric strength.  The third static term is due to \textit{uncorrectable modes} in the AO system.   The image sharpening techniques we utilize, described in Section 2.7, cannot completely correct the modes introduced by the PPMs themselves, although the low-order components are canceled nearly completely by the woofer static correction.   As the $83\%$ static Strehl (at $658$ nm) is uniform across the field, this error term has no spatial dependence. 

The fourth static term is due to minor \textit{misalignments in the field stops}.  These field stops have to be removed every AO iteration for checking Strehl and put back into place for wavefront sensing.  The value quoted ($10$ nm) is derived from Fresnel simulations of wavefronts passing through spatial filters; if the holes are displaced by a minor amount, the overall wavefront is filtered differently and changed.  This error term is dependent on the repeatibility of the filter wheels that mount the focal plane stops.  They are estimated to be repeatable to the $100\; \mu$m level, while the size of the holes are $1.6$ mm.  

The fifth and sixth static terms are \textit{linearity calibration drift} and \textit{WFS zeropoint drift}.  The zeropoint error is determined by comparing measured wavefronts with references saved at a fixed time in the past.  Over the typical length of AO simulations (2 hours), the errors increase by $\sim10$ nm.  The linearity calibration drift error is determined by measuring the open-loop accuracy of the wavefront sensors on long timescales.  This is done by comparing the multiple LGS wavefronts on a common ground layer.  We find that this open-loop accuracy degrades by $\sim10$ nm over 2 hours.  It is assumed that both the WFS zeropoint references and the linearity calibration data can be re-taken before every experiment.  Both of these drift terms are due to optical drift, isolated to several non-kinematic cube beamsplitter mounts leading to the Shack-Hartmann WFS facility.

\subsubsection{Time-dependent Terms}
The following error terms have a dependence on total atmospheric strength.  The \textit{atmospheric fitting error} is given by \citep{har98}
$$\sigma_{fit} = 0.53\; (d/r_0)^{(5/6)}$$
where $d$ is the subaperture size and $r_0$ is the Fried parameter.  The total atmospheric strength is given by 
$$\sigma_a = 0.366 \;(D/r_0)^{(5/6)}$$
where $D$ is the telescope diameter.  Solving for $\sigma_{fit}$ for our system (66 subapertures across a 10-meter pupil) gives
$$\sigma_{fit} = 0.0441\; \sigma_a.$$
Similarly, the \textit{WFS aliasing error} can be expressed as a fraction of the fitting error ($40\%$ for Shack-Hartmann WFS) and thus the atmospheric error \citep{fus06}.  $$\sigma_{alias} = 0.0176\; \sigma_a$$
The error in estimating the wavefront in open loop is labeled the \textit{WFS systematic} error.  This error is reduced from $\sim80$ nm to $\sim30$ nm through linearity calibration \citep{amm07}.   The open-loop error model is characterized by measuring the disagreement between adjacent laser guide stars on common ground layer turbulent plates.  The WFS Systematic term is empirically measured to be proportional to the total atmospheric strength with the following fitted dependence:
$$\sigma_{WFS} = 0.045\; \sigma_a.$$  
The \textit{scintillation term} is calculated explicitly from theoretical models of scintillation \citep{str94}.  The scintillation index is given by 
$$\sigma_\chi^2 = 0.288 \;(\sqrt{\lambda h} / r_{0, layer})^{(5/3)}$$
where $\lambda$ is the testbed wavelength, $h$ is the height of a particular atmospheric layer, and $r_{0,layer}$ is the Fried parameter of that layer.   The effect on the measured Strehl is 
$$S = exp(-\sigma_\phi^2)exp(-\sigma_\chi^2).$$
Substituting our physical path lengths, we can express this error as a function of the atmospheric strengths at different altitudes:
$$\sigma_\chi = 1.466 \;(\lambda / a^2)^{(5/12)} \sqrt{h_1^{(5/6)}\sigma_{a,1}^2 + h_2^{(5/6)}\sigma_{a,2}^2}$$
where $a$ is the physical telescope aperture size in meters, $h_x$ is the height at layer $x$ in meters, and $\sigma_{a,x}$ is the atmospheric strength at that altitude.  This error can be explicitly calculated at each simulation timepoint, as the relative strength at each altitude is a derived output of the tomographic algorithm (TSW).  

The \textit{WFS Scintillation} error is the effect that scintillation has on Shack-Hartmann wavefront sensors.  Ideally, wavefront sensors are not sensitive to amplitude variations, but in reality this is not true.  We explicitly measure this error by comparing the wavefronts measured on a single plate when at different altitudes.  We assume that this error has the same dependencies as pure scintillation:  $\sigma_{\chi, wfs} \;\propto\; h^{(5/12)}$ and $\sigma_{\chi, wfs} \;\propto\; \sigma_{a,x}.$  The final model includes both of these proportionalities and is calibrated via the empirical measurement above.  This model is:
$$\sigma_{\chi, wfs} = 0.235\;  \sqrt{h_1^{(5/6)}\sigma_{a,1}^2 + h_2^{(5/6)}\sigma_{a,2}^2}$$
Note that both pure scintillation and WFS scintillation are severe for this testbed, as the physical aperture size is much smaller compared to the wavelength of light than a 10-meter telescope.  

Finally, notice that the Strehl slowly drops over the course of four realizations in Figure 6.  This is due to a slow drift in pupil-to-lenslet registration, amounting to $0.5\%$ of a subaperture per real hour.  The simulations were taken over 2 days, so this amounts to a pupil shift of $25\%$ of a subaperture ($0.38\%$ of the full aperture, or $68 \;\mu$m physical shift at the telescope pupil).  Using simulations of shifted Kolmogorov atmospheres, we find that this results in $\sim 40$ nm of error after 2 days.  Then
$$\sigma_{pupil} = 0.89 \times t,$$ 
where the units of $\sigma_{pupil}$ are in nanometers and $t$ is the number of hours since pupil calibration.  As the PPM pixels are 5 times smaller than the Hartmann subapertures, slow pupil drift is not a problem for software and control, as long as it can be calibrated frequently.  Note that closed-loop AO systems would suffer a great deal from this magnitude of pupil shift (perhaps not converging) because wavefronts from previous AO iterations are erroneously shifted multiple times; a DM-to-lenslet registration with accuracy better than 5\% of a subaperture is typically desired for closed-loop systems.  Fortunately, open-loop AO systems have no feedback, so the only error introduced is that modeled by the equation above.

In addition to a standard DM-to-lenslet calibration involving raising and centroiding 25 Gaussians on the PPM, we also perform a spatially-variant, Strehl-optimal calibration to lock down the pupil drift.  This involves placing a layer of turbulence at the ground, measuring the mean wavefront, and varying the x- and y-shift of the applied wavefront until the Strehl on a particular science star is maximized.  When performed on all seven science test stars individually, this surprisingly produces a set of offset vectors that extend radially outward from the central star, implying that the PPM and the ground layer are not perfectly conjugated.  The magnitude of the pupil drift error ($\sigma_{pupil}$) is measured directly by repeating this calibration over long time baselines.  

\subsubsection{Tomographic Error}
The only error term dependent on both space and time is the tomographic error.  Tomographic estimation error is due to a number of contributors, including misestimation of layer heights, inadequate coverage of atmospheric space, and tomographic blind modes.  In these experiments, we fix layer heights and ensure that no atmosphere is missed, so the dominant component is that due to tomographic blind modes.  Tomographic blind modes are three-dimensional modes that are unsensed due to the geometry of the LGS constellation.  It is expected that this error is directly proportional to atmospheric strength.  

Tomographic error increases with off-axis distance, as less information is available to estimate the wavefront.  We calibrate this dependency explicitly with numerical simulations of atmospheric tomography, using the same solver used to analyze bench data (TSW).  We match the LGS constellation parameters and Kolmogorov atmospheric strengths in these simulations to those in the laboratory experiments.  This gives tomographic errors at different field points that correspond to individual atmospheric realizations.  The averages of these values are shown in Table 3.

\section{CONCLUSIONS}  

\subsection{Error Budget Meta-Analysis}

The error budget terms describing the MOAO/LTAO bench can be grouped into three categories:  Traditional errors, calibration errors, and dynamic drift errors.   The ``traditional'' error terms are those endemic to adaptive optics, limited by subaperture size and laser power, that cannot be improved with better hardware calibration.  These include fitting error, wavefront sensor aliasing, tomographic error, scintillation, and photon error.  The ``calibration'' errors include those terms that could nominally be reduced with increased hardware cost and calibration effort, including WFS systematic error, deformable mirror go-to error, static uncorrectable errors, and wavefront sensor scintillation.   The ``dynamic drift'' errors include those calibration errors that have a time dependence, requiring a greater frequency of calibration.  These include linearity calibration drift, pupil registration drift, WFS zeropoint drift, and field stop misalignment.  The total magnitudes of the errors in each of these categories is illustrated in Figure 11.

\begin{figure}
   \begin{center}
   \begin{tabular}{c}
   \includegraphics[height=6cm]{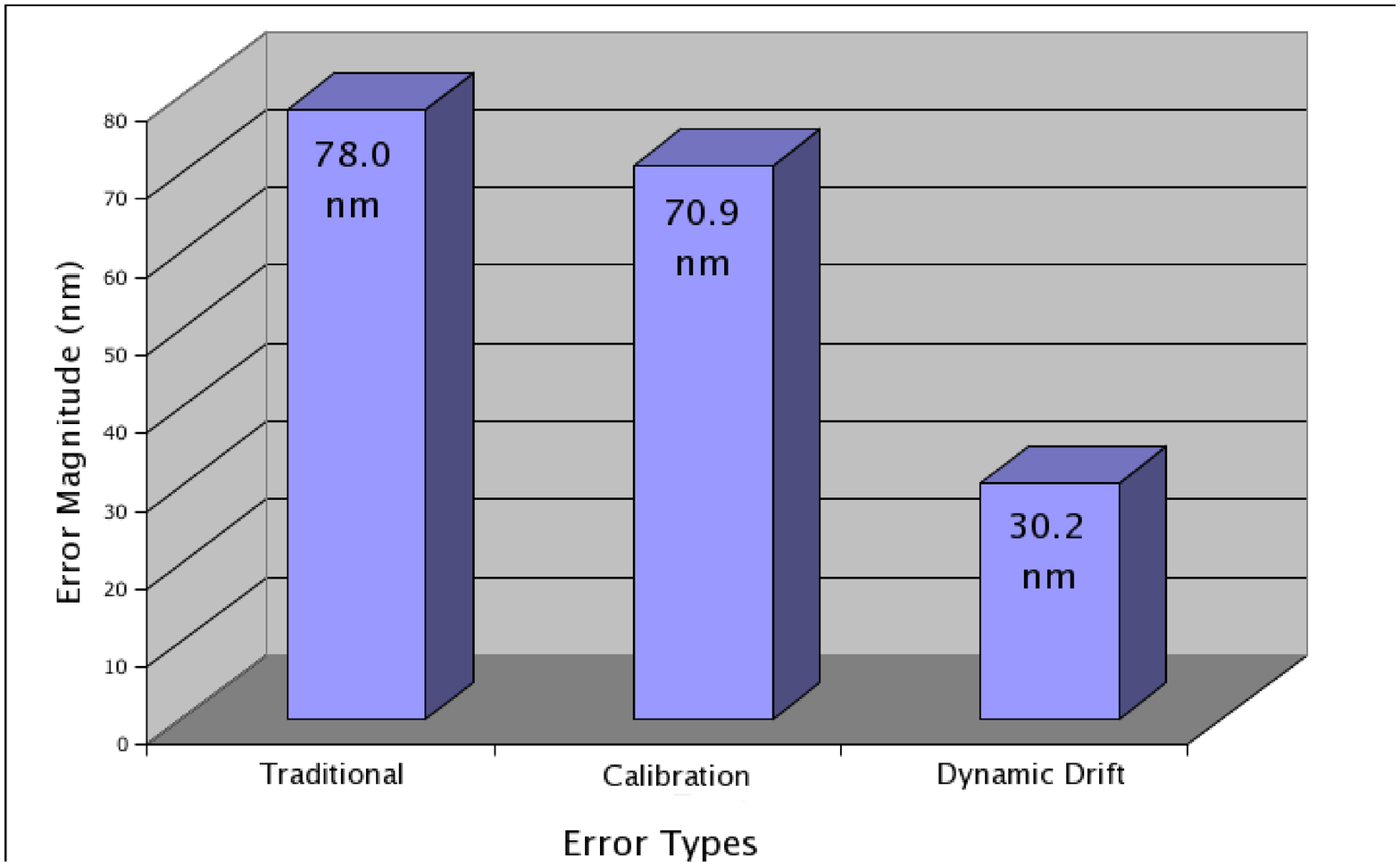}
   \end{tabular}
   \end{center}
   \caption[example] 
{ \label{fig:error_budget_summary} 
Error Budget Summary for on-axis case in laboratory units of nanometers.  The error terms are grouped into three categories, with total values shown.}
   \end{figure} 

The total magnitude of calibration errors is comparable to the traditional AO errors, but the dynamic drift component has been greatly reduced by extensive, frequent calibration.  In quadrature, the magnitude of all calibration errors is 49\% of the total error budget for the on-axis case, with 42\% arising from static errors.  The remaining contributor of 51\% is from traditional error terms that cannot be reduced without decreasing subaperture size and increasing laser power.  This \textit{partitioning of errors}, with static and dynamic calibration errors comparable to traditional error terms, is expected for high-order AO systems such as the MOAO/LTAO testbed.  

The vibration amplitude on this testbed is much lower than that seen in typical telescope instruments, but the magnitude of pure optical drift is larger, which requires optical calibration every few hours.  These calibrations include (1) dark frame measurement, (2) a Hartmann linearity calibration to enable open-loop wavefront sensor performance,  (3) a DM-to-lenslet registration procedure to calibrate high-order optical warping (for both PPM and woofer), and (4) spatially-variant image sharpening to increase static Strehl throughout the field of view.  We also use a single ground-layer turbulent plate to optimize (1) open-loop gain, (2) WFS registration between LGS pupils, and (3) global DM-to-lenslet registration as a function of science star position.  

It will be important to include extensive calibration facilities when designing visible-light LTAO systems for routine use at astronomical observatories.  Of particular utility will be accelerometers on all powered optics, which can be used to diagnose the low- and high-order effects of telescope and instrument vibration.  Full telescope simulators, including LGS elongation mechanisms and 3D turbulators, will assist in daytime calibration.  If the go-to characteristics of the deformable mirrors change with time, it may be necessary to periodically transfer them to accurate interferometers external to the instrument (``external deformable mirror ports'') for recalibration.

\subsection{Future Testbed Plans}
The coming decade brings exciting new possibilities for high-order adaptive optics on large telescopes.  Infrared AO system designs for Extremely Large Telescopes (ELTs) and specialized high-contrast architectures on 8-10 meter telescopes \citep[GPI,][]{mac06} have required the development of order 64x64 deformable mirrors.   As these mirrors have been functionally demonstrated and even higher order DMs are being designed, there exists the potential to lower subaperture sizes on observatory AO systems to those suitable for visible-light AO correction.   We have demonstrated in this paper that order $64\times64$ for a 10 meter telescope may be sufficient to realize visible-light correction in the bright tip/tilt star case, given a site with good median seeing (0\farcs65).  However, increasing LGS power in step with decreasing subaperture size remains an expensive option.  For wide field correction in the visible, the tomographic error due to blind modes quickly becomes the dominant error term on 6-10 meter telescopes.  We intend to investigate one remedy to the tomographic error that does not sacrifice field size or require an order of magnitude more laser guide stars:  Wind prediction.  We have performed preliminary experiments with wind prediction on the MOAO/LTAO testbed and found that it assuages the build-up of tomographic error in sparsely-sampled rings in upper-altitude metapupils, provided the atmosphere is 100\% Taylor frozen flow \citep{amm08}.

We believe the MOAO/LTAO experiments presented here demonstrate technologies necessary for wide-field, laser-driven, visible-light AO with high sky coverage.  These include WFS calibrations that enable open-loop operation, open-loop deformable mirror characterization, and pupil registration calibration.  

New technologies like uplink LGS correction may enable the power densities necessary for visible-light AO correction \citep{gav08}.  Hartmann centroiding accuracy for a laser guide star is proportional to the inverse square of its angular size, so measurement error can be greatly reduced by pre-correcting the outgoing laser beam and ``sharpening'' its image on the sodium layer.  Moderate sharpening of the LGS spot to Hartmann pixel sizes ($0\farcs8 - 1\farcs0$) preserves the use of Hartmann sensors, but delivers 4-5 times the effective power of the unsharpened LGS (typically $2\farcs0$).  Uplink correction is being tested on-sky with the Visible Light Laser Guide Star Experimental System \citep[ViLLaGEs,][]{gav08}.

\section{ACKNOWLEDGEMENTS}	
This work has been supported in part by the NSF Science and Technology Center for Adaptive Optics, managed by the University of California (UC) at Santa Cruz under the cooperative agreement No. AST-9876783.  

The MOAO/LTAO testbench in the Laboratory for Adaptive Optics at UCSC has been funded by the Gordon \& Betty Moore Foundation.  S.M.A acknowledges Bachmann fellowship support by the Allen family through UC Observatories/Lick Observatory, the UCSC Graduate Division, and the National Science Foundation through the G.R.F program.  E.A.L. acknowledges support through the UC Riverside Chancellor's Fellowship Program.  

\facility{}

{}

\end{document}